\documentclass[conference]{IEEEtran}
\usepackage{amssymb}
\usepackage{amsmath}
\usepackage{mathtools}
\usepackage{spconf,amsmath,epsfig}
\usepackage{algorithm}
\usepackage{algorithmic}
\usepackage{graphicx, subfigure}
\usepackage{graphicx}
\usepackage{url}

\begin{document}

\title{On Modeling Geometric Joint Sink Mobility with\\ Delay-tolerant Cluster-less Wireless Sensor Networks}

\name{M. Akbar$^{\pounds}$, N. Javaid$^{\pounds}$, A. A. Khan$^{\pounds}$, Z. A. Khan$^{\$}$, U. Qasim$^{\S}$}

\address{$^{\pounds}$COMSATS Institute of Information Technology, Islamabad, Pakistan. \\
        $^{\$}$Faculty of Engineering, Dalhousie University, Halifax, Canada.\\
        $^{\S}$University of Alberta, Alberta, Canada.}
\maketitle

\begin{abstract}
Moving Sink (MS) in Wireless Sensor Networks (WSNs) has appeared as a blessing because it collects data directly from the nodes where the concept of relay nodes is becomes obsolete. There are, however, a few challenges to be taken care of, like data delay tolerance and trajectory of MS which is NP-hard. In our proposed scheme, we divide the square field in small squares. Middle point of the partitioned area is the sojourn location of the sink, and nodes around MS are in its transmission range, which send directly the sensed data in a delay-tolerant fashion.  Two sinks are moving simultaneously; one inside and having four sojourn locations and other in outer trajectory having twelve sojourn locations. Introduction of the joint mobility enhances network life and ultimately throughput. As the MS comes under the NP-hard problem, we convert it into a geometric problem and define it as, Geometric Sink Movement (GSM). A set of linear programming equations has also been given in support of GSM which prolongs network life time.
 \end{abstract}
\IEEEpeerreviewmaketitle

\section{Introduction}
WSNs are based on tiny sensors, and are used for sensing the data which we can not calculate directly, as the decisions is based on the accuracy and availability of the information. Fields where, direct collection of data is impossible, sensors are used to sense the data and give the complete set of required information. Its popularity is increasing with every passing day, as these networks are useful in many scenarios, such as, battle field, environment monitoring, soil moisture sensing and body area networks, etc. The only constraint with wireless sensors is that, they have limited power due to their size, for this reason they work in small bandwidths. Nodes once deployed in a field are hard to track and move, like if they are thrown in battle field through plan, physically their tracking is not possible. So, recharging and replacement of the batteries is difficult. Nodes have to transmit sensed data to the sink, located anywhere inside or outside the field. Traditional network follow multi hop approach in which nodes act as a relay for other nodes and drain their energies quickly. When nodes start depleting their energies, instability enters in the network. New wave of research is study of WSNs with MS, so that the resources with the energy constrained can save the energy and the lifetime can be increased.

\section{Related Work}
In literature, researchers are working hard, in terms of introducing different approaches/ schemes to make WSNs efficient. In past, protocols were designed with static sink but now MS is changing the direction of research. MS is independent of energy constraint, and can save the energy which CHs consume in aggregation and transmission of member nodes. There are two types of mobilities, controlled mobility \cite {1}, in which sink  moves on predefined path. Other is random mobility \cite {2}, where, sink moves randomly in the whole field and collects the data. In literature, different authors are tackling mobility in different ways and trying to over come challenges \cite {3}, \cite {4}, \cite {5}, \cite {6} and \cite {7}. In \cite {8} author exploited the sink mobility in delay-tolerant WSN, \cite {7} MSRP solves controlled movement of sink and avoid the hole problem and thus increases lifetime of the network. In \cite {9} and \cite {10}, authors are considering single hop transmission with a single MS with time scheduling and reducing energy consumption of sensors in relaying data.

We are going to compare our model with the clustering schemes. Lot of work had been done on different clustering schemes, initially in Energy-efficient communication protocol for wireless micro sensor networks (LEACH) \cite {11}. In this paper author selected a homogeneous network in terms of energy and applied the clustering scheme in which field was divided into small clusters and every cluster has a Cluster Head (CH) which is selected on the basis of energy. Nodes are supposed to send the data to the CH and CH forward it to the sink. Following this scheme \cite {12}, author selected two level heterogenous network on the basis of energy and performed the same operation and get improved results. In above mentioned both schemes threshold is defined for nodes both, advanced and normal, which decides the criteria of election of CH in every epoch. Every node has knowledge of network energy and it uses election probability and become CH on the basis of remaining energy. When nodes reach at the threshold, sink can check the heterogeneity of the network.

\section{Motivation}
Introducing mobility with the sink increases network life time. The load between nodes is distributed and every node directly transmits data to the sink. MS is considered as a vehicle which is able to refill fuel or recharge its battery. Issues with a MS are different as compared to the static sink. Looking for any optimal sink trajectory is related to NP-hard \cite {13}. Here, for the simplicity it is converted into a geometric sink model. Speed of the sink is selected intuitively, as it should not be very fast that nodes may not be able to transfer complete information and also not very slow that nodes which are waiting for sink start overflowing the data. MS is passing through the trajectory and harvesting the data from the nodes.

In \cite{12}, two level heterogenous network was proposed for gathering data through clustering. Energy is consumed in cluster formation and in electing CHs. Also, CH acts as a relay for the other CHs which are far from sensing range of the sink, otherwise if they follow direct transmission then the CH which are far from sink will have to use double transmission energy. In our model we have considered two level network consisting on normal and advance nodes where deployment of nodes is random. Two sinks move on the pre-defined pattern and gather data. To save the energy we introduced sleep and awake modes; when MS enters in sensing range of a sensor they are awake and transmit the data to the sink and if sink is out of range they stay in sleep mode.

\section{Network Model}
WSN is modeled as directed graph $G=(\mathcal{V}, \mathcal{E})$, where $\mathcal{V}=V\cup S$. $|V|=N$, set of wireless sensor nodes, $S$ is the set of sinks with $|S|=K$. For all $i, j \in \mathcal{V}$, $\exists(i, j)\in \mathcal{E}$ if and only if $i$ and $j$ are within a square transmission range $r_{tx}$.
\\- All the sinks are moving at the same time means, their movement is synchronous. Sink movement is equally distributed on all locations in every epoch $n = 0, 1, 2, ....$ All the data is transferred to sink before it further proceed to next location, so, there is no data is pending between epochs. During node to sink, data delivery is representing by $C_j$, and it is defined as $C_j \in {0, 1}$, $C_j = 1$ if sink sojourn location is $v_j$; otherwise $C_j = 0$, $1 \leq j \leq n.$
\\- $x_{ij}\in {0, 1} $, when sink is at location $v_j$, then $x_{ij} = 1$, where $v_j, 0\leq i,j \leq n.$ $x_{ij}$ is the rate of information.
\\- The sensor nodes are two level heterogeneous stationary network. Initial energy is $E_0$ for normal nodes and for advanced nodes it is $(1+ \alpha)E_0$, it is $\alpha$ times greater than normal nodes. We are considering energy consumption only in transmission and reception of the information.
\\- Network lifetime is define here as, the time interval until first node dies. When the energy of node is deplete, node dies and after the death of first nodes instability period of the network starts.
\\By using above assumptions we are now formulating the optimization problem of finding maximum lifetime using mobile sinks on different trajectories with delay tolerant behaviour.
\begin{eqnarray}
     Maximize T = \sum_m t_m    
     \end{eqnarray}
subject to:
   \begin{subequations}
\begin{equation}
\\ \sum_{s\in S} C_j(n)=1 \qquad \forall j\in V     
\end{equation}
\begin{equation}
\\ \sum_{e^-} f_i(n) - \sum_{e^+}f_i(n)= X \qquad \forall i \in V           
\end{equation}
\begin{equation}
\\ \sum_{m} t_m (\sum_{i\in N} E_{ij}^{tm} + \sum_{k\in N} E_{ki}^{rm}) \leq E_{initial} \qquad \forall i,j,k \in N           
\end{equation}
\begin{equation}
\\ x_{ij}-R_{ij}\leq 0 \qquad \forall i,j           
\end{equation}
\begin{equation}
\\ x_{ij},\quad t_i\geq 0 \qquad \forall i,j
  \end{equation}
\end{subequations}
where, $e^-= \{e \in E\mid e= (v, n), n \in V\}$ and $e^-= \{e \in E\mid e= (n, v), n \in V\}$. The function $f_i$ is the amount of data send over an edge during epoch $n$. In equation (1) defines the objective function which is maximization of lifetime. Equation (2a) shows that in one epoch only one sink is locating at a node,equation (2b) is a flow constraint, $X$ is the total packets in epoch $n$, which is difference between the received and transmitted flow. Equation (2c) is the energy conservation equation. Equation (2d) is the rate constraint which is explaining that total information rate which is flowing through the link $(i, j)$ should not exceed the link capacity $R_{ij}$ which is the upper bound of the transmission rate.

\subsection{GSM}
Movement of the sink on predefined trajectory belongs to NP- hard problem. To simplify this problem, we have converted it into simple geometric case, GSM.
Under consideration network is scalable, and assumption is that field will be remain square. Main objective for all the assumptions is to maximize the network  life time and introducing delay tolerant characteristic in the field. We have divided field in such a way that all nodes find minimum distance from the sink, and sinks are mobile without having energy constraint. $d_i = |s_k - NUC (v)|$, shows that $d_i$ is the Euclidean distance between sink $s_k$ and Node Under Consideration (NUC). Here, $s \in S$ and $v \in V$ during epoch $n$. According to basic network calculus delay bound $D_i$ is defined as \cite {14}:
\begin{align}
 D_i&=h(\overline{\alpha}_i),\beta_i)\\
 &=Sup_{s\geq 0}\{inf \{\tau \geq0: \overline{\alpha}_i(s)\leq \beta_i (s+ \tau)\}\}\\
 \overline{D}_i&=\sum_{i\in v} D_i\\
 \overline{D}&= max_{i=1,...,N} \overline{D}_i.
\end{align}
Where, $\overline{\alpha}_i$ is the sensed input and $\beta_i$ is service curve. This analysis is for FIFO scheduling at the sensor nodes, which is use in the common practical cases.

\begin{figure}[ht]
\centering
\includegraphics [height=9 cm,width= 8 cm]{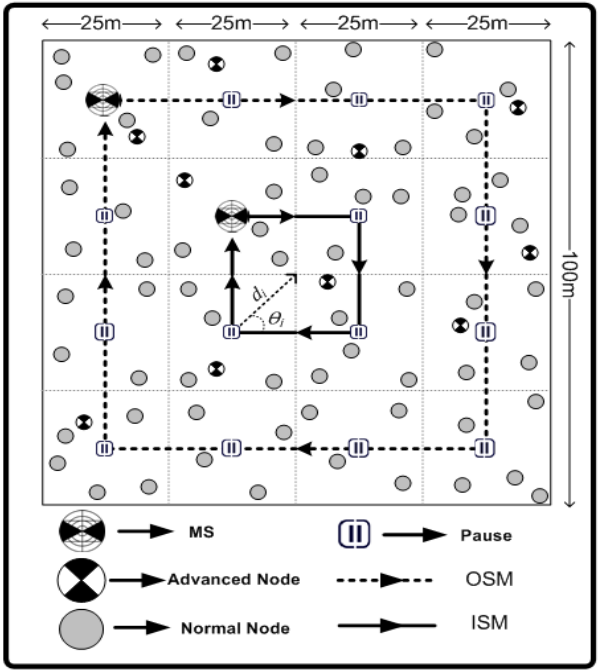}
\vspace{-0.3cm}
\caption{Network Topology}
\end{figure}

In our model we divided square field into small square regions, shown in fig 1. There are sixteen small squares in a considered square field, twelve out of sixteen are lying on the outer boundary and four are inside this boundary. In the outer boundary, we have considered separate area in which one sink is gathering data while second sink is supposed to collect the data from inner region. These two sinks start their journey simultaneously for the data collection from the respective regions where, the trajectories of sinks are square too. One sink is collecting data from inner squares and other is from outer squares. We predefined the trajectories and the sink locations, in every square; sink will stop in the middle of it will receive data from the nodes directly. When sink moves forward from the current location the nodes lying in that small square will observe sleep mode and the nodes which come under the new stop region will be awake and start transmission. Only those nodes will be active in both trajectories which are present in the portioned region where MS is present. The transmission range is set accordingly maximum distance a node can have from the sink is $d_i = \sqrt{2}x$, and the distance between two sojourn locations is $d_{min} = 2x$ and two different trajectories sojourn location is $d_{max}=2\sqrt{2}x$, as shown in fig. 2.

\begin{figure}[ht]
\centering
\includegraphics [height=6 cm,width= 7 cm]{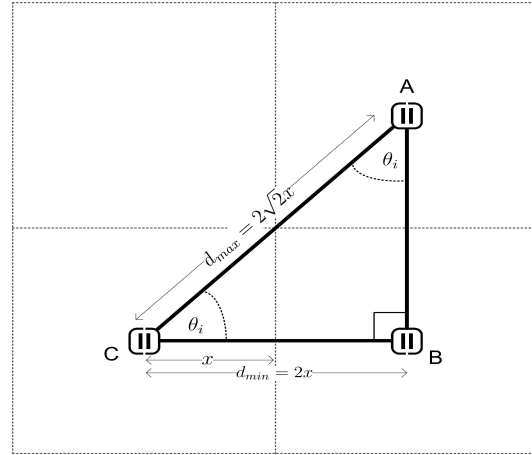}
\vspace{-0.3cm}
\caption{Depicting $d_{max}$ and $d_{min}$}
\end{figure}

\section{Simulation Results}
We perform our simulations in a WSN with the dimensions of $100m \times 100m$. Total number of nodes deployed in a network are $N = 100$, in which $10\%$ are advanced and remaining are normal nodes. Placement of the nodes in the field is random, it means both horizontal and vertical axes are selected randomly within the upper and lower values of network. We are considering joint sinks, one is moving in inner square trajectory, and second is in outer square trajectory. The maximum distance of the nodes with the MS related to their partitioned squares is $d_{min}$. Initially energy assigned to the normal node is $0.5 Joules$ and in advanced nodes is $1 joule$. We have compared our simulation results with SEP and LEACH.
\begin{figure}[ht]
\centering
\includegraphics [height=7.25 cm,width=9.25 cm]{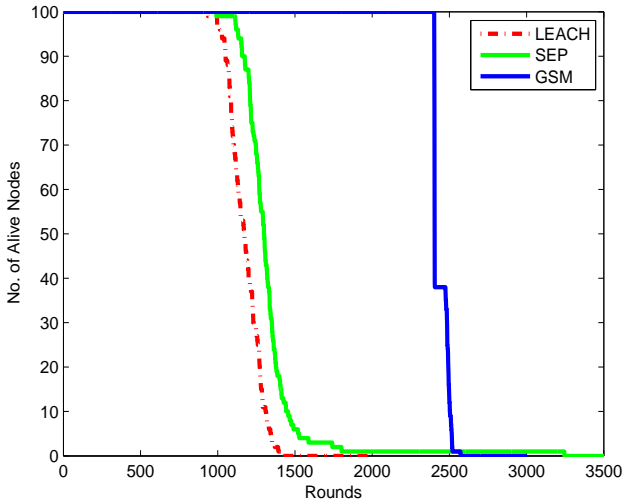}
\vspace{-0.3cm}
\caption{Number of alive nodes per round}
\end{figure}

In fig. 2 we have compared the results of our proposed scheme GSM with the SEP and LEACH. First node of the LEACH runs out in First node of the SEP depleted in 1000 rounds and after that within few rounds nodes will be dead, almost in 1500 rounds whole network is dead. Now look at the SEP its performance is better than the LEACH, its first node dies at 1150th round and after that normal nodes die in few more rounds the advanced nodes keep alive the system upto 2000th round. SEP is performing better than LEACH as it has 10\% advanced nodes, and these nodes get maximum chance of becoming CH. Energy is utilized in aggregation, reception and transmission. Due to advanced nodes SEP is performing good. Now comparing SEP with our proposed GSM, it is also virtually divided in small regions with a defined geometry. CHs are replaced by MSs, here, energy consumption of aggregation and election of CH is saved. As the whole network is efficiently divided into regions and MS observes fixed path and nodes are delay-tolerant, this results in saving energy and prolonging the network life. As first node dies in 2400th round and last in 2600th, stability period is enhanced as compared to previous schemes. After the death of the first node, remaining nodes deplete quickly as compared to others.
  \begin{figure}[ht]
\centering
\includegraphics [height=7.25 cm,width=9.25 cm]{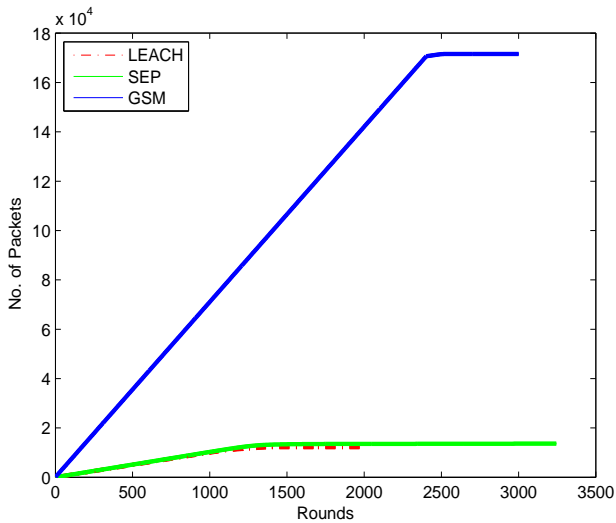}
\vspace{-0.3cm}
\caption{Throughput of the network}
\end{figure}

In fig. 3 results of the throughput are compared. Throughput is defined as total data sent to the sink, sensed by sensors. Heterogeneity increases the stability and ultimately throughput. It also depends upon the link capacity; if all nodes send data at a time, a bottle neck will be created and throughput will decrease, as there are chances of data loss. In joint square sinks throughput is increased because nodes in the network observe sleep and awake mode. As the field is divided in to the small squares, when MS stops at any sojourn location the related square of that stop become awake and starts sending sensed data when MS stops at any sojourn location. Due to staying in sleep mode when MS is busy in collecting data from the far stops, nodes minimize their energy consumption. Also nodes are saving energy by not acting as a relay for data transmission and CH selection.  Throughput of LEACH and SEP  is close to each other as SEP is extended by adding 10\% advanced nodes. After the death of first node, network enters in the instability region and remains constant. Throughput of the joint square sink is significantly greater in stable and as well as in instable region.
\begin{figure}[ht]
\centering
\includegraphics [height=7.25 cm,width=9.25 cm]{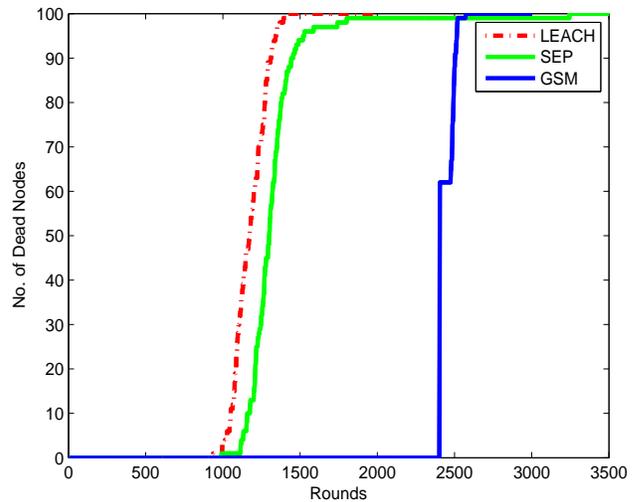}
\vspace{-0.3cm}
\caption{No. of dead nodes per round}
\end{figure}
Plots for the dead nodes are shown in fig. 4. Network lifetime is defined as the time span in which all the nodes are alive. When first node is dead an instable period starts and after few rounds whole network becomes dead. The main emphasis is on load balancing and energy reservation. In given plots LEACH enters in the instable period earlier as compared to SEP. Our joint sinks is performing much better than both. Load is well balanced between the nodes which are also saving their energy while in sleep mode.
Above figures show that our proposed scheme of joint sink mobility performs significantly better than conventional clustering schemes. As we have used the concept of joint sink mobility in GSM, in which two sinks are moving inside the sensing field and gathering the sensed data from all deployed sensor nodes. We have defined sensing range of MSs, so, all the nodes within the vicinity of the MSs transmit their data if they lie in the sensing range of that sink. Otherwise nodes go into sleep mode and save energy. GSM is only applicable for delay tolerant applications; both of the sinks are moving in a squared trajectory within the sensing field with different dimensions. In clustering schemes, sensor nodes send data to their associated CHs and then CHs further transmit the aggregated data to BS. Whereas, in GSM MS have replaced the CHs which save energy of the nodes, as MS is independent of energy constraint. In LEACH and SEP there is instability region, after the death of first node, it takes few hundreds of round more till the last node dies, but in GSM in a very less time whole network dies thus providing greater stability period.
\section{Conclusion}
In proposed GSM, the network model is two level heterogenous and also observes the nodes both in sleep and awake modes. Field is virtually divided into small squares so that we can point the sink location and also calculate maximum distance of the MS from the nodes residing in the square. By considering these factors, we have enhanced throughput and ultimately network life by considering these factors. We compared GSM with LEACH and SEP, both are supporting clustering. We observe that in LEACH nodes are homogenous and die quickly while in SEP stability period increases due to advanced nodes, stability period increases. In GSM sink motion is in a controlled pattern, but more random experiments can be planned so that the motion of the sink is controlled accordingly. Also, number of sinks can be increased.  In future, our goals are to explore more trajectories and geometries of sink to minimize the transmission cost and prolong the network life.

\end{document}